\documentclass[10pt,twocolumn]{article}

\usepackage{graphicx}
\usepackage{amsmath,amssymb}
\usepackage[table]{xcolor}
\usepackage{colortbl}
\usepackage{bm}
\usepackage{bbm}
\usepackage{empheq}
\usepackage{fancyhdr}
\usepackage{makecell}
\usepackage{arydshln}
\usepackage{tabularx}
\usepackage{cite}
\usepackage{diagbox}
\usepackage{multirow}
\usepackage{subcaption}
\usepackage{hhline}
\usepackage{chemfig}
\usepackage{xcolor}
\usepackage{comment}
\usepackage{etoolbox}
\usepackage{multirow}
\usepackage{array}
\usepackage{makecell}
\usepackage[version=4]{mhchem}
\AtBeginEnvironment{tabular}{\setlength{\minrowclearance}{2pt}}

\definecolor{TableHeaderColor}{RGB}{220, 225, 230} 
\definecolor{TableCellColor}{RGB}{245, 247, 250}   

\definecolor{HeaderDark}{RGB}{30, 50, 75}
\definecolor{RowTint}{RGB}{240, 245, 250}

\setchemfig{
atom sep=2.2em,
bond style={line width=0.6pt},
double bond sep=2pt
}

\usepackage[
    a4paper,
    margin=1.8cm,
    columnsep=0.7cm
]{geometry}

\definecolor{primary}{RGB}{0,139,139}      
\definecolor{secondary}{RGB}{180,40,40}    
\definecolor{accent}{RGB}{40,120,120}      
\definecolor{header}{RGB}{40,65,90}
\definecolor{lightrow}{RGB}{235,242,248}

\definecolor{HeaderOrange}{HTML}{8C3A00}   

\definecolor{LightOrange}{HTML}{FDF2EB}    

\usepackage{hyperref} 
\hypersetup{
    colorlinks=true,
    linkcolor=primary,
    citecolor=accent,
    urlcolor=secondary
}

\usepackage{titlesec}

\titleformat{\section}
  {\normalfont\large\bfseries}
  {\thesection.}
  {0.5em}
  {}

\titleformat{\subsection}
  {\normalfont\normalsize\bfseries}
  {\thesubsection.}
  {0.5em}
  {}

\begin{document}

\twocolumn[
\begin{center}

{\LARGE\bfseries Quantum chaotic signature in\\[6pt] non-empirical ensemble of DNA base pairs}

\vspace{0.4cm}

Mehdi Ameri

\vspace{0.2cm}

{\itshape
Department of Physics, Faculty of Science, Ferdowsi University of Mashhad\\
P.O. Box 1436, Mashhad, Iran\\[4pt]
\normalfont\small Email: \texttt{amerimehdi9755@gmail.com}
}

\vspace{0.4cm}

\end{center}

\begin{abstract}
The statistical study of many-body chaotic quantum systems would indicate important phenomena such as short-range level repulsion and long-range spectral rigidity. We investigate such behavior in an ensemble of gas-phase guanine-cytosine (GC) and adenine-thymine (AT) Watson-Crick (WC) base pairs by calculating their Kohn-Sham molecular orbital energy levels via ab initio density functional theory (DFT). We track the probability distribution, adjacent-gap ratios (AGR), number variance, and spectral compressibility of the unfolded level spacings to see whether their statistics follow the statistics of the eigenvalues of random matrices. After comparing the results to Poisson, semi-Poisson (SP), and Gaussian Orthogonal Ensemble (GOE) statistics, it becomes evident that the statistics of orbital energy levels agree closely with the semi-Poisson distribution, pointing to an intermediate regime between integrability and chaos, characterized by level repulsion and strong long-range correlations. Potential biophysical implications for charge transfer and related biological processes are also discussed.
\end{abstract}

\vspace{0.3cm}


\vspace{0.6cm}
]


\section{Introduction}
The molecular structure of DNA operates at a scale where quantum mechanics plays a crucial role in shaping its dynamics and symmetries. The double-helical structure of a DNA sequence consists of molecular complexes held together by interactions governed by molecular orbitals, corresponding to a minimum on the potential energy surface (an optimized geometry) under fixed environmental conditions such as temperature and solvent effects. Altering these conditions drives the system out of equilibrium, inducing a dynamical trajectory toward stabilization. A fundamental question arises: how chaotic is this trajectory? Addressing this requires interpreting chaos in quantum systems. Substantial progress has been made in this direction \cite{BGS-conj,Mehta1,Mehta2,int-RMT}, showing that quantum chaotic systems exhibit three signature features in their spectrum.


First, the energy-level spacing distribution of such a system matches the eigenvalue spacing distribution obtained from an ensemble average of random matrices. A random $\mathrm{N\times N}$ matrix is a matrix whose entries are drawn from random variables. In the simplest case, these variables can be independent and identically distributed. Independence means that the value of one matrix entry does not influence or determine the value of any other entry, while identical distribution means that all entries are drawn from the same probability distribution, sharing the same probability density function, mean, and variance. However, this does not hold for all random matrices. In this work, we consider entries drawn from Gaussian distributions that, depending on their symmetry class (real symmetric, complex Hermitian, or quaternion self-dual), can introduce correlations among entries and need not be identically distributed.

Second, nearest-neighbor energy levels repel as if they experience a Coulomb force, keeping the spectrum discrete without degeneracy. This feature is sometimes referred to as {\it short-range spectral rigidity}.

\begin{figure}[!htbp]
    \centering
    \begin{subfigure}[b]{0.3\textwidth}
        \centering
        \includegraphics[width=\textwidth]{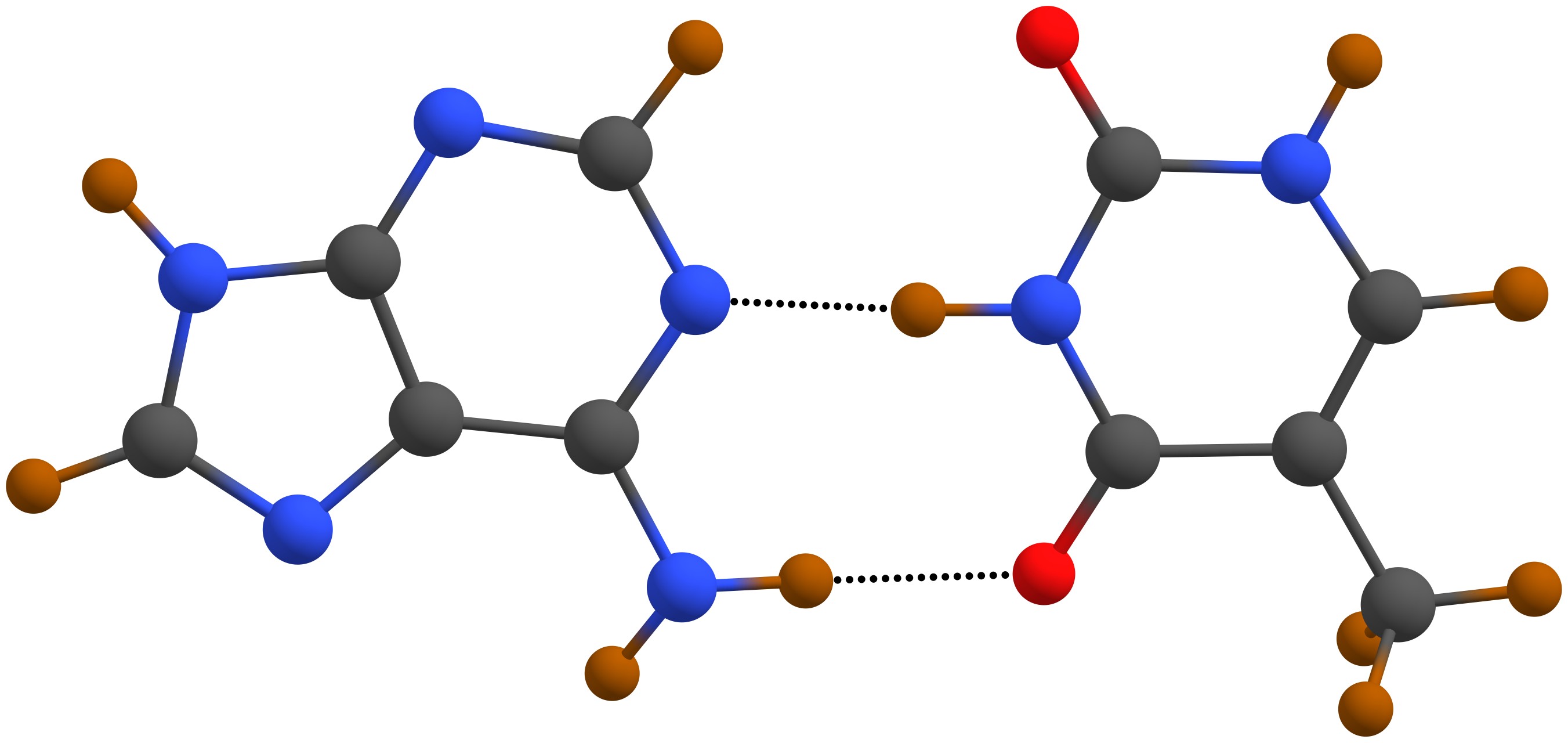}
        \subcaption*{Watson-Crick AT} 
    \end{subfigure}
    
    \vspace{1em}
    
    \begin{subfigure}[b]{0.3\textwidth}
        \centering
        \includegraphics[width=\textwidth]{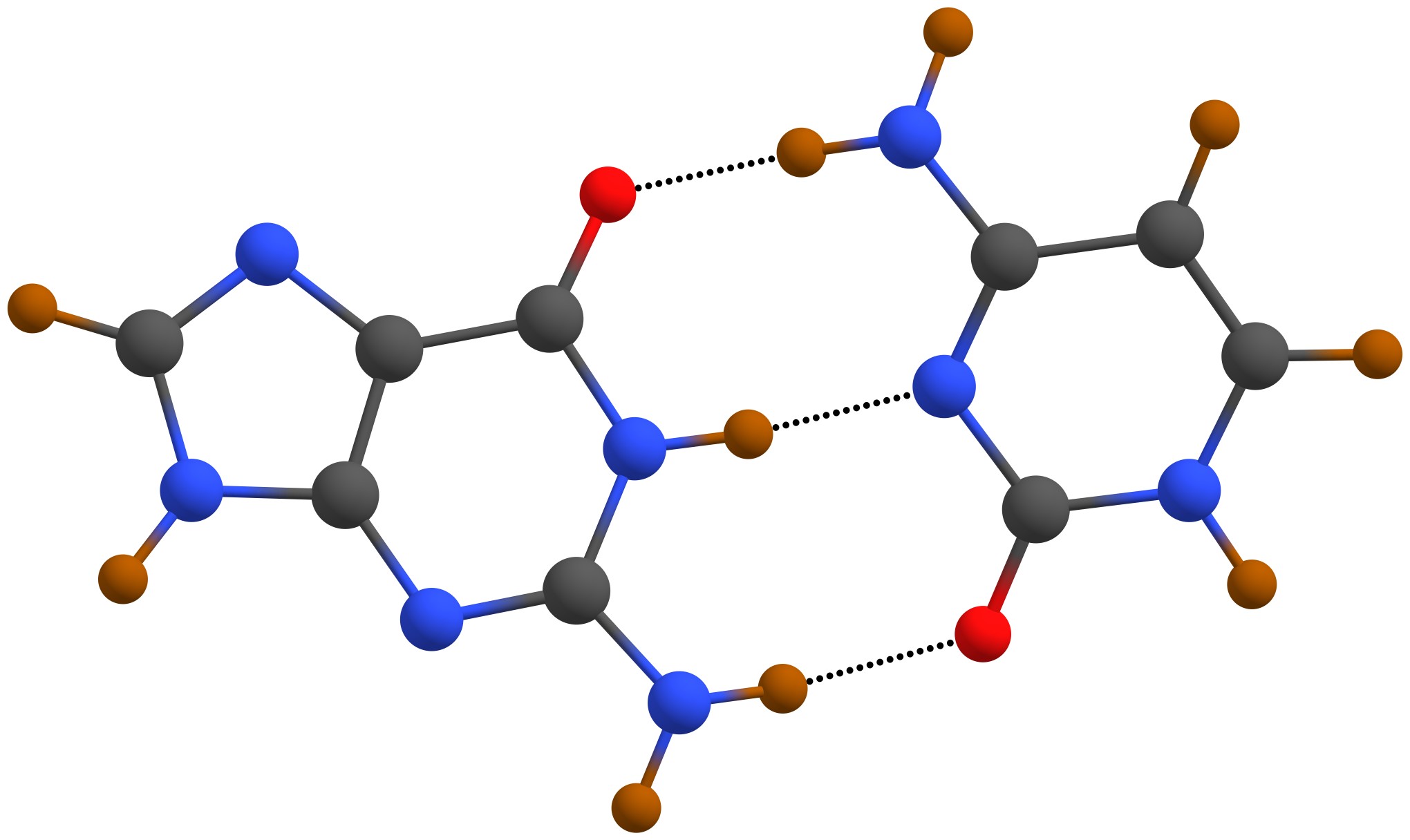}
        \subcaption*{Watson-Crick GC}
    \end{subfigure}
    \hspace{0.3cm} 

    \caption{Optimized geometry of Watson-Crick base pairs with zero net charge and spin multiplicity. Black, red, blue, and brown spheres represent carbon, oxygen, nitrogen, and hydrogen atoms, respectively. Covalent and non-covalent (hydrogen) bonds are represented by solid and dotted lines, respectively. In AT, the $\mathrm{R(N\cdots H-N)}$ and $\mathrm{R(N-H\cdots O)}$ distances are $1.78$ and $1.90\text{ \AA}$, respectively. In GC, the $\mathrm{R(O\cdots H-N)}$, $\mathrm{R(N-H\cdots N)}$, and $\mathrm{R(N-H\cdots O)}$ distances are $1.81$, $1.78$, and $1.89\text{ \AA}$, respectively.}
    \label{fig:opt-bps}
\end{figure}

Third, distant energy levels are correlated via an effectively logarithmic interaction, producing a chain-like rigidity across the spectrum. This feature is commonly referred to as {\it long-range spectral rigidity}. \\
A completely different line of analysis reveals a striking correlation between nearest-neighbor base pairs (bps) \cite{Luo1998,Albuquerque2005,Knoch2009}. A notable study of DNA random walks likewise suggests long-range correlations among intron-less genes \cite{Peng1992}. Although these studies rely on distinct approaches, they point toward a possible universal explanation for base-pair correlations.\\
One can therefore examine the level-spacing statistics and base-pair correlations in their molecular electronic properties - such as orbital energies - to determine whether short and long-range correlations exist among the energy levels, which in turn helps answer our earlier question.
\\
Therefore, one needs to calculate the electronic energy levels of a large ensemble of base pairs and compare their level-spacing distribution with those from random matrix theory. To this end, we generate an ensemble of 1,000 isolated Watson-Crick GC and AT base pairs by randomly distorting the optimized geometry (most stable configuration) of each complex shown in Fig.~\eqref{fig:opt-bps} in two ways: (1) hydrogen-bond variation and (2) multimode variation, detailed in Sec.~\eqref{HB} and Sec.~\eqref{MM}, respectively. Watson-Crick base pairing describes the specific hydrogen-bonding interactions between complementary DNA bases \cite{WC}: adenine pairs with thymine via two hydrogen bonds, whereas guanine pairs with cytosine via three. These purine-pyrimidine pairs stabilize the antiparallel DNA double helix and enable the faithful storage and replication of genetic information.

The most favorable interaction energy is obtained from geometry optimization, which yields a single value corresponding to a minimum on the potential energy surface. Consequently, distorting the optimized geometry produces less stable or non-equilibrium configurations. These distortions can be viewed as small atomic vibrations or other perturbations acting on the base pair over time.\\
The generated ensemble consists solely of distinct base-pair geometries at $0~\mathrm{K}$, omitting backbone strands, solvent, and other external effects. Because this is a simulated set of base-pair geometries, we refer to it as a non-empirical ensemble describing gas-phase base pairs. We use density functional theory to compute the electronic properties of this ensemble, a topic discussed in the following subsection.

\subsection{Density functional theory}
Density functional theory is a quantum-mechanical method used to investigate the electronic structure and properties of many-electron systems, particularly molecular complexes.
The theoretical foundation of DFT lies in the Hohenberg-Kohn theorems \cite{HK}, which establish that the ground-state properties of an interacting many-electron system are {\it uniquely} determined by its electron density $\rho(\mathbf{r})$ rather than its many-electron wavefunction. Here $\mathbf{r}=(x,y,z)$ denotes the position vector in three-dimensional space, and $\rho(\mathbf{r})$ is the electron density at $\mathbf{r}$, describing how electrons are spatially distributed around the nuclei.
\\ On the other hand, the Kohn--Sham equation \cite{KS}:
\begin{align}
\left[-\frac{\hbar^2}{2m}\nabla^2+V_{\mathrm{eff}}(\mathbf r)
\right]\phi_i(\mathbf r)=\varepsilon_i\phi_i(\mathbf r)\,,
\end{align}
practically maps the interacting many-electron problem onto an auxiliary system of non-interacting electrons moving in an effective potential:
\begin{align}
V_{\mathrm{eff}}(\mathbf r)=V_{\mathrm{ext}}(\mathbf r)+V_{\mathrm{H}}(\mathbf r)+V_{\mathrm{xc}}(\mathbf r)\,,
\end{align}
where $m$ is the electron mass, $\phi_i(\mathbf r)$ are the Kohn-Sham single-particle orbitals (interpreted as molecular orbitals) and $\varepsilon_i$ are their corresponding eigenvalues. The external potential $V_{\mathrm{ext}}(\mathbf r)$ describes the Coulomb attraction between the electrons and the nuclei, the Hartree potential $V_{\mathrm{H}}(\mathbf r)$ represents the classical electrostatic repulsion of the electron density, and the exchange-correlation potential $V_{\mathrm{xc}}(\mathbf r)$ accounts for quantum-mechanical exchange (arising from fermionic antisymmetry and the Pauli exclusion principle) as well as electron correlation. The latter effect arises because mutual Coulomb repulsion correlates the instantaneous motions of electrons, lowering the probability of finding two electrons in close proximity. The electron density enters the effective potential through $V_{\mathrm{H}}(\mathbf r)$ and $V_{\mathrm{xc}}(\mathbf r)$ via:
\begin{align}
V_{\mathrm H}(\mathbf r)=\frac{e^2}{4\pi\epsilon_0}\int\frac{\rho(\mathbf r')}{|\mathbf r-\mathbf r'|}\,d^3r'\,,~
V_{\mathrm{xc}}(\mathbf r)=\frac{\delta E_{\mathrm{xc}}[\rho]}{\delta \rho(\mathbf r)}\,,
\end{align}
where $e$ is the elementary charge and $\epsilon_0$ is the vacuum permittivity. Here $E_{\mathrm{xc}}[\rho]$ is the exchange-correlation energy, expressed as a functional of the electron density. The density itself is constructed from the occupied Kohn-Sham orbitals via:
\begin{align}
\rho(\mathbf r)=\sum_{i}^{\mathrm{n}}f_i|\phi_i(\mathbf r)|^2\,,
\end{align}
for $n$-paritcle system with occupation numbers $f_i$. Now to obtain the effective potential, and consequently the obrital eigenvalues, one needs to apply the self-consistent procedure as follows: (1) Start with an initial guess of mathematical basis-functions for electron density, (2) Construct the effective potential, (3) Solve the Kohn-Sham equations to obtain the orbitals and eigenvalues, (4) Calculate a new electron density from the obtained orbitals, (5) Repeat until the electron density and total energy converge.
This iterative feedback mechanism is why DFT calculations are called self-consistent field (SCF) calculations.
\\ The exact exchange-correlation energy functional is not generally known; therefore, practical DFT calculations employ suitable approximations. For our calculations, we use the B3LYP hybrid functional together with the def2-SVP basis set, a widely used combination in computational quantum chemistry \cite{B3LYP}. We carry out these calculations using the PySCF package \cite{pyscf} in Python~3\footnote{All scripts and data files are available at this \href{https://doi.org/10.5281/zenodo.24522824}{webpage}.}, obtaining the interaction energies, ground-state energies, and - most importantly - the molecular orbital energies across the ensemble of randomized AT and GC base-pair geometries.

It has been widely shown in the literature that electrostatic interactions contribute significantly to interactions between molecular systems, with additional contributions arising from subsystem polarizability and dispersion (van der Waals) interactions. Moreover, one must account for basis-set superposition error (BSSE) when computing interaction energies with finite basis sets. BSSE is an artificial overestimation of the attraction between two subsystems that occurs because, when they come into close proximity, each fragment can borrow basis functions from the other to lower its energy.
\\The interaction energy quantifies how energetically favorable it is for two subsystems A and B to interact:
\begin{align}
E_{\text{int}}=E_{\mathrm{AB}}-(E_{\mathrm{A}}+E_{\mathrm{B}})+\mathrm{BSSE}\,,
\end{align}
where $E_{\mathrm{AB}}$ is the total ground-state energy of the combined complex $\mathrm{AB}$, and $E_{\mathrm{A}}$ and $E_{\mathrm{B}}$ are the single-point energies of the isolated monomers $\mathrm{A}$ and $\mathrm{B}$ (typically evaluated at their geometries in the complex).
\\
Interaction and ground-state energies do not exhibit quantum chaotic behavior for the following reasons. First, the interaction energy is a difference between total energies rather than an eigenvalue of a quantum Hamiltonian. Second, across an ensemble of AT or GC configurations, the ground-state energy fluctuates around an average value and therefore does not provide a rich enough spectrum of values for meaningful level statistics. Consequently, we focus on the orbital energies, which are eigenvalues of the Kohn-Sham Hamiltonian $\hat{H}_{\mathrm{KS}}\phi_i=\varepsilon_i\phi_i$. Because these eigenvalues vary noticeably across distorted geometries, we analyze their spacing and distribution statistics for the two ensembles generated via HB and MM variations.

Theoretical calculations of DFT are useful because they allow controlled comparisons. You can change one factor at a time - for example, replace a base, alter a substituent, remove a hydrogen bond, or compare different stacking arrangements - without needing to synthesize every possible molecule experimentally.
At the same time, practical DFT studies typically rely on simplified models, whereas native DNA exists in an aqueous environment involving thermal motion, counterions, explicit solvent molecules, broad conformational ensembles, and entropic contributions. Incorporating all of these realistic effects explicitly is possible in principle, but incurs a tremendous computational cost in both CPU time and memory.

\subsection{Spectral statistics}
Suppose we have an ordered sequence of energy eigenvalues $\lambda_1<\lambda_2<\cdots<\lambda_{\mathrm{qN}}$ for $q$ realizations of $\mathrm{N}\times \mathrm{N}$ random matricies. The nearest-neighbor level spacings are defined as
\begin{align}
s_i=\lambda_{i+1}-\lambda_{i}\,.
\end{align}
Integrable systems typically follow the Poisson distribution,
\begin{align}
P(s)=e^{-s}\,,
\end{align}
while the quantum chaotic systems follows a rather different pattern given by Wigner-Dyson disterbution
\begin{align}
P_{\mathrm{WD}}(s)=as^{\beta}e^{-bs^2}\,.
\end{align}
The constants $a$ and $b$ are chosen so that $P_{\mathrm{WD}}(s)$ is normalized to unity and has unit mean spacing, $\langle s \rangle = 1$. The parameter $\beta$ is the Dyson index, whose value reflects the system's symmetry class and takes $\beta = 1, 2,$ and $4$ for the Gaussian Orthogonal Ensemble (GOE; real symmetric matrices, $\mathrm{M}_{ij} = \mathrm{M}_{ji}$), Gaussian Unitary Ensemble (GUE; complex Hermitian matrices, $\mathrm{M}_{ij} = \mathrm{M}_{ji}^{*}$), and Gaussian Symplectic Ensemble (GSE; real quaternion self-dual matrices; see \cite{dyson}), respectively. Among these canonical ensembles, we focus here on the $\beta = 1$ GOE case, for which $a = \pi/2$ and $b = \pi/4$.\\
In addition, intermediate systems - neither conventionally integrable nor fully quantum chaotic - often display the semi-Poisson distribution studied in \cite{SP},
\begin{align}
P_{\mathrm{SP}}(s)=4se^{-2s}\,.
\end{align}
At first glance, we see that near $s=0$, the Poisson peaks at unity, which shows no sign of repulsion between eigenvalues while $P_{\mathrm{WD}}(s)$ and $P_{\mathrm{SP}}(s)$ are both proportional to $s$, which indicate very small gaps are suppressed. Plus, all these functions decays to zero for large $s$, indcating that large spacings are increasingly unlikely. However, nearest-neighbor spacing distributions alone do not fully capture correlations between nearby and distant levels. To probe short-range level repulsion and long-range spectral rigidity more directly, we turn to two complementary diagnostics: adjacent-gap ratios and the spectral number variance, which probe short and long-range correlations, respectively.\\
Adjacent-gap ratios ($r$) are a spectral statistics metric used to analyze energy level-spacing fluctuations in quantum systems and random matrix theory. Instead of looking at individual level spacings in isolation, they compare the sizes of consecutive energy gaps along a sorted spectrum. The adjacent-gap ratio for consecutive spacings $s_i$ and $s_{i+1}$ is commonly defined as:
\begin{align}
r_i = \frac{\min(s_i, s_{i+1})}{\max(s_i, s_{i+1})}\,.
\end{align}
Because $0 \le \min(s_i, s_{i+1}) \le \max(s_i, s_{i+1})$, every ratio $r_i$ lies strictly within the interval $[0, 1]$. It is a suitbale probe for short-range correlations and level repulsion because in local comparison it involves three neighboring energy levels ($\lambda_i, \lambda_{i+1}, \lambda_{i+2}$), so it directly reflects how nearby levels cluster or push each other apart.\\
For Poisson disterbtuition, nearby levels are independent and can come arbitrarily close to one another, so small spacings $s_i \to 0$ occur frequently. Consequently, $r_i$ retains substantial weight near $0$ with a mean value of $\langle r \rangle \approx 0.386$, reflecting the absence of level repulsion.
\\Level repulsion in the GOE pushes nearby levels apart and suppresses very small spacings. Because $\min(s_i, s_{i+1})$ rarely becomes small compared with $\max(s_i, s_{i+1})$, the probability density near $r \to 0$ is strongly suppressed and the distribution shifts toward larger ratios, yielding a mean value of $\langle r \rangle \approx 0.536$.
\\Suppression near $r \to 0$ is present in semi-Poisson indicating real repulsion between nearby levels, but weaker than GOE, placing the mean in $\langle r \rangle \approx 0.50$.
\\ Because $r_i$ relies only on consecutive spacings, it does not capture correlations between energy levels separated by larger intervals along the spectrum. To probe long-range correlations and spectral rigidity, we evaluate the number variance, defined as
\begin{align}
\Sigma^2(L)=\left\langle [\mathcal{N}(L)-L]^2 \right\rangle\,,
\end{align}
where $L$ is the length of an interval in the unfolded spectrum (normalized to unit mean level density) and $\mathcal{N}(L)$ is the number of energy levels inside that interval. The average $\langle \cdots \rangle$ runs over window positions across the spectrum and across all ensemble realizations, quantifying how strongly the eigenvalue count inside an interval of mean length $L$ fluctuates from window to window. For large-$L$, the number variance of the afromentioned benchmarks has the following number variance asymptotic limit\footnote{Although we also checked the exact finite-$L$ expressions \cite{dysonIV}, the differences from the asymptotic forms are minor; we therefore use the asymptotic expressions here for clarity.}\cite{SP}:
\begin{align}
&\Sigma_{\mathrm{P}}^2(L)\sim L\,,~~~
\Sigma_{\mathrm{SP}}^2(L)\sim \frac{L}{2}+\frac{1}{8}\,,\\[6pt]
&\Sigma_{\mathrm{GOE}}^2(L)\sim\frac{2}{\pi^2}
\left[\ln(2\pi L)+\gamma+1-\frac{\pi^2}{8}\right]\,,
\end{align}
where $\gamma\approx0.57721$ is the Euler-Mascheroni constant. The spectral compressibility defined as
\begin{align}
\chi=\lim_{L\rightarrow\infty}
\frac{\Sigma^2(L)}{L}
\end{align}
provides a compact measure of long-range correlations in the energy spectrum by characterizing the asymptotic growth rate of eigenvalue-count fluctuations within an interval of unfolded length $L$. It is a dimensionless quantity bounded between $\chi = 0$ for rigid spectra in the GOE (where $\Sigma^2(L)$ grows only logarithmically with $L$) and $\chi = 1$ for uncorrelated levels obeying Poisson statistics (where $\Sigma^2(L) \sim L$). Intermediate semi-Poisson statistics sit halfway between these limits with $\chi = 1/2$.

We use the mean value of the adjacent-gap ratio $\langle r\rangle$, the scaling behavior of the number variance $\Sigma^2(L)$, and the spectral compressibility $\chi$ as primary metrics to probe correlations in orbital energy level spacings and to assess whether the spectrum aligns with the Poisson, semi-Poisson, or GOE random matrix benchmarks. The numerical results and comparisons are detailed in sec.\eqref{HB} and sec.\eqref{MM} for the hydrogen-bond and multi-mode structural variations, respectively.

To ensure reliable comparisons against random matrix benchmarks, careful attention must be paid to spectral {\it unfolding}. To examine universal statistical properties of energy-level spectra-here, the Kohn-Sham orbital energies - one first removes the influence of a non-uniform mean density of states, rescaling the spectrum so that the mean level density becomes approximately uniform (typically unit density). This separates genuine spectral fluctuations and level correlations from smooth, system-specific variations in the average density of states. In our analysis, we estimate a common spectral density from the pooled orbital eigenvalues across all geometries in the ensemble using a Gaussian kernel density estimator (KDE) \cite{KDE}, with the bandwidth selected automatically via Silverman's rule of thumb \cite{Silverman}.\\
The resulting cumulative density of states is used as a smooth unfolding mapping applied individually to the orbital spectrum of each distorted geometry. The nearest-neighbor spacings $s_i$ are then calculated directly as the differences between consecutive unfolded energy levels.\\
Finally, the unfolded spacings are normalized by their geometric mean within each sample, enforcing a unit mean spacing $\langle s \rangle = 1$ so that level-spacing distributions across different distorted configurations can be consistently pooled and compared on a uniform scale.
\\To reduce possible edge effects in the unfolding procedure, the lowest and highest 10$\%$ of the energy levels are discarded.





\section{Hydrogen-bond length variation\label{HB}}

We begin by evaluating the orbital energy statistics under a single-variable geometric perturbation: freezing the internal molecular structures of the individual nucleobases while systematically varying only the intermolecular hydrogen-bond distances in AT and GC base pairs. Specifically, the varied parameters are the hydrogen-bond lengths $R(\mathrm{X\text{--}H}\cdots\mathrm{Y})$ and $R(\mathrm{X}\cdots\mathrm{H\text{--}Y})$, where $\mathrm{X}$ and $\mathrm{Y}$ denote donor and acceptor heavy atoms involved in covalent (solid line) and hydrogen-bonded (dotted line) contacts. Note that $\mathrm{R}$ here represent only the hydrogen bond length.

In the fully relaxed, optimized Watson--Crick (WC) configurations, these hydrogen bonds occupy potential energy minima corresponding to equilibrium bond lengths $R_{\mathrm{eq}}$. To sample non-equilibrium and less stable configurations, we randomly perturb $R$ within a bounded interval of $[1.50\,\text{\AA}, 3.00\,\text{\AA}]$. The lower limit of $1.50\,\text{\AA}$ prevents unphysical steric overlap and severe electron cloud repulsion between donor and acceptor atoms, whereas the upper cutoff of $3.00\,\text{\AA}$ excludes non-interacting, fully decoupled fragments.

Starting from the Cartesian coordinates of the optimized AT and GC base pairs (Fig.~\ref{fig:opt-bps}), we randomly shift the relative positions of the rigid monomers along the hydrogen-bonding vectors to generate ensemble populations of 1,000 distinct distorted geometries for both AT and GC complexes. Finally, we execute single-point DFT calculations across these 1,000-geometry ensembles to extract the corresponding Kohn--Sham molecular orbital spectra.

Because single-point DFT calculations across 1,000 perturbed base-pair configurations require substantial computational resources, evaluating the complete Kohn--Sham spectrum for every geometry is computationally demanding. However, we can achieve significant speedups by recognizing the structural constraints of the rigid-monomer scan: because the internal bond lengths and angles within each constituent nucleobase remain strictly frozen, the intra-monomer electronic structure is largely invariant. Consequently, each isolated monomer can be approximated as maintaining a fixed single-point background, allowing the spectral variations of the dimer complex to be evaluated efficiently by focusing predominantly on the distance-dependent intermolecular overlap across the hydrogen-bonding interface.
\\
Across the 1000-geometry ensemble, each single-point calculation for each geometry yields 321 molecular orbitals (69 occupied and 252 virtual) for the AT complex and 316 molecular orbitals (68 occupied and 248 virtual) for the GC complex\footnote{The total orbital count reflects the basis-set expansion dimension ($N_{\mathrm{basis}} = N_{\mathrm{occ}} + N_{\mathrm{virt}}$) under the def2-SVP discretization.}. Across all perturbed configurations, the full energy spectrum spans from $-520.099\,\mathrm{eV}$ to $+106.009\,\mathrm{eV}$ for AT, and from $-520.325\,\mathrm{eV}$ to $+105.866\,\mathrm{eV}$ for GC.

Following the KDE spectral density estimation and boundary trimming protocol, we unfold the central 80$\%$ of orbital energy levels across all 1000 perturbed configurations. This procedure yields an ensemble total of 256,000 normalized level spacings for the AT base pair and 253,000 normalized spacings for the GC base pair.
Finally, we unfold the orbital energy levels and then obtain the normlized level-spacings with total number of 256,000 for AT and 253,000 for GC base pairs.  Across the pooled dataset, the normalized level spacings $s_i$ range from a minimum of $4.87 \times 10^{-4}$ to a maximum of $12.39$ for AT, and from $1.52 \times 10^{-3}$ to $9.94$ for GC.

\begin{figure}[!htbp]
    \centering
    
    \begin{subfigure}[b]{0.4\textwidth}
        \centering
        \includegraphics[width=\textwidth]{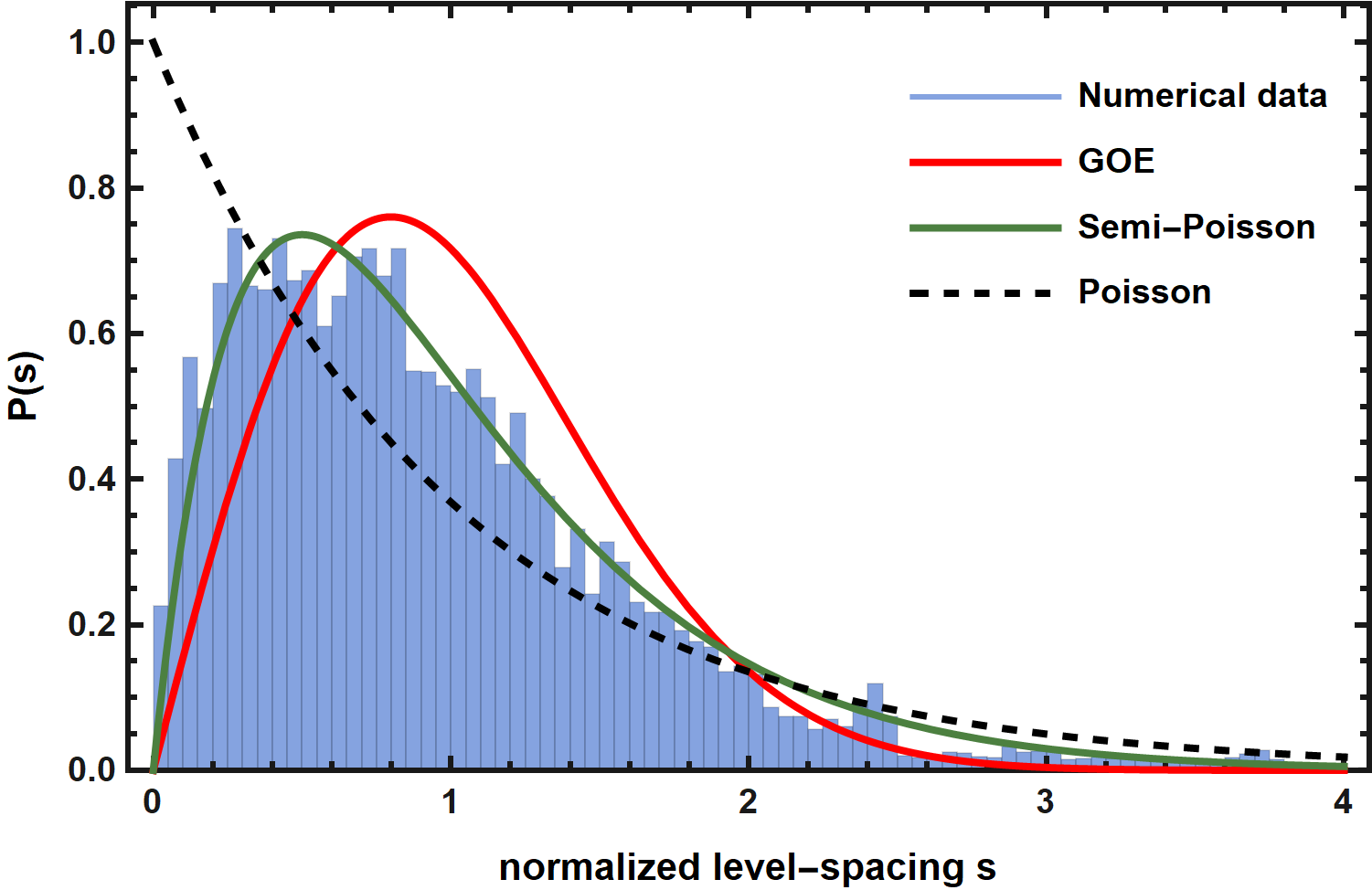}
        \subcaption*{(a) level-spacings of orbital energies for AT.}
    \end{subfigure}
    
    \vspace{1em}
    
        \begin{subfigure}[b]{0.4\textwidth}
        \centering
        \includegraphics[width=\textwidth]{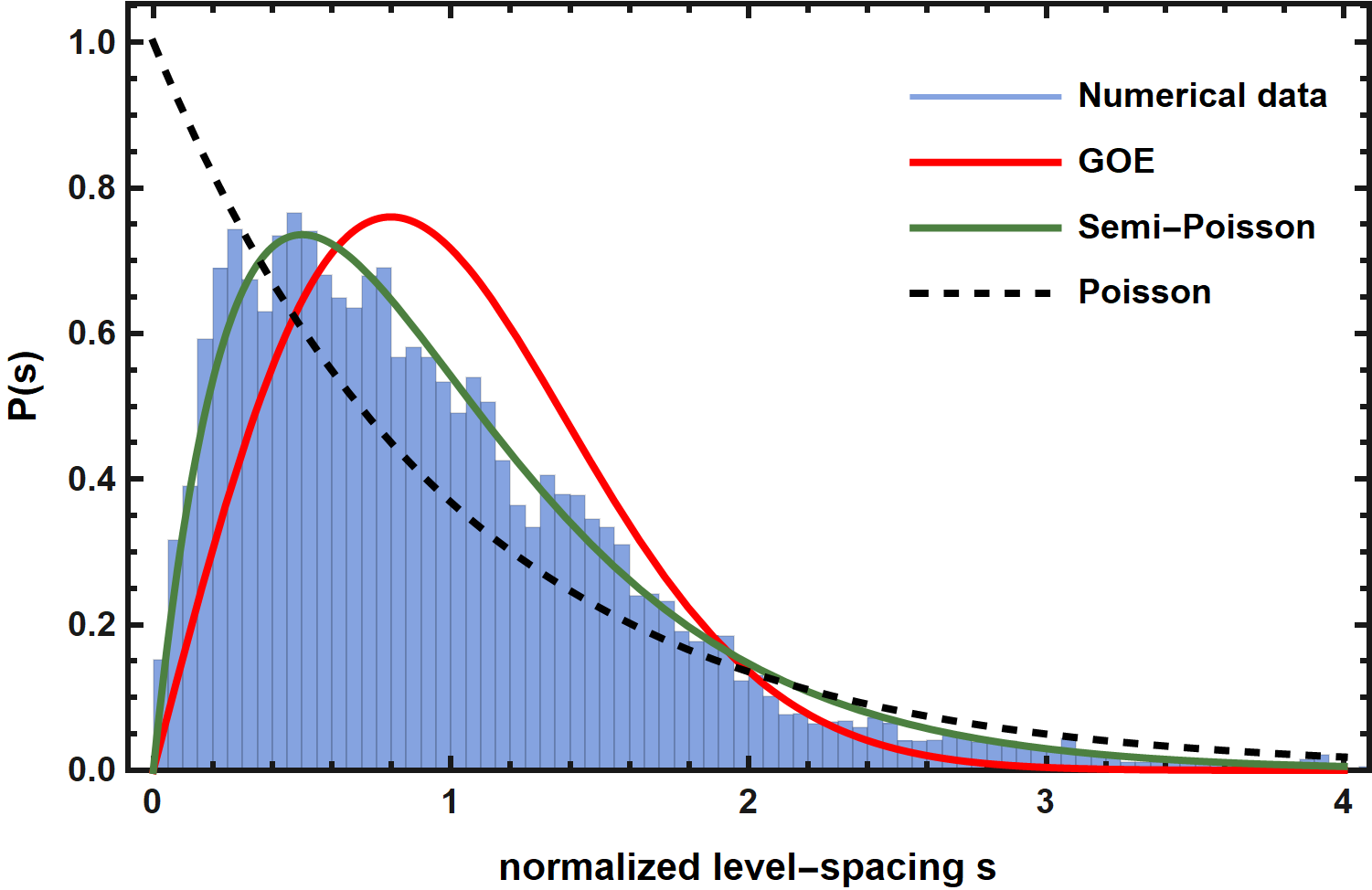}
        \subcaption*{(b) level-spacings of orbital energies for GC.} 
    \end{subfigure}
    
    \vspace{1em}
    
    \begin{subfigure}[b]{0.4\textwidth}
        \centering
        \includegraphics[width=\textwidth]{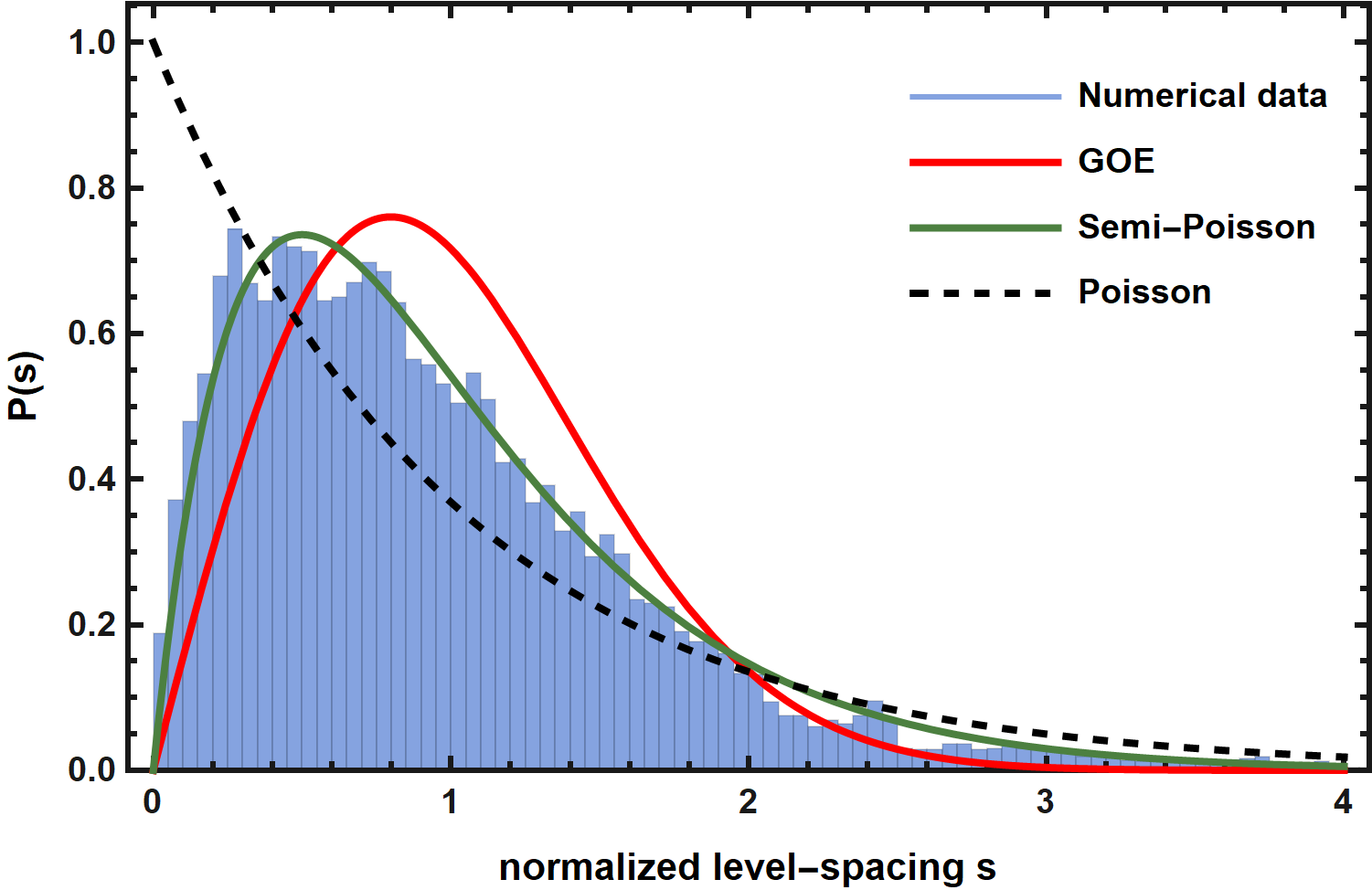}
        \subcaption*{(c) level-spacings of orbital energies for both bps.}
    \end{subfigure}
    \hspace{0.3cm} 

    \caption{The three plots show the probability distribution of normalized level-spacings of orbital energies for HB variation, evaluated for AT, GC, and combined base pairs with a bin width of $\Delta s = 0.05$. Theoretical curves are drawn for comparison.}
    \label{fig:OE-HB}
\end{figure}

The resulting level-spacing distributions $P(s)$ for the hydrogen-bond length variation are presented in fig.\eqref{fig:OE-HB} for the AT base pair, the GC base pair, and the pooled dataset combining both complexes. Across all three distributions, the numerical data show distinct oscillations that closely trace the semi-Poisson profile, while displaying a clear deviation from the standard GOE.

Although these level-spacing distributions clearly demonstrate eigenvalue repulsion and suggest long-range spectral correlations, $P(s)$ alone is insufficient to quantify the precise strength of these correlations. To obtain a definitive diagnostic, we evaluate the adjacent-gap ratios and the spectral number variance across the unfolded orbital spectra.

As detailed in table.\eqref{tab:AGR-Chi}, the adjacent-gap ratios yield an average value of $\langle r \rangle = 0.465$ across all three analysis sets (AT, GC, and combined), falling very close to the theoretical semi-Poisson benchmark of $\langle r \rangle_{\mathrm{SP}} \approx 0.50$. This confirms that nearest-neighbor orbital energy levels in the hydrogen-bond perturbed ensemble experience real level repulsion, operating in a regime intermediate between Poisson and GOE. The number variance also reveals a consistent signature, as discussed below.

To calculate the number variance, unlike individual spectral unfolding, in which a separate density function is estimated for each realization, in the present procedure we employ a single common KDE unfolding transformation for the entire ensemble.\\This uniform transformation preserves overall energy scale consistency and global level density across all configurations, preventing artificial suppression or enhancement of long-range level number fluctuations across window intervals of length $L$.

After unfolding, the energy levels of each molecular geometry $j$ are sorted in ascending order, yielding the unfolded spectrum $\{\lambda_i^{(j)}\}$. For a given interval length $L$, we consider sliding windows of the form $[x, x + L)$, where $x$ denotes the starting position of the interval. The number of unfolded energy levels contained within such an interval for the $j$-th realization is defined as $\mathcal{N}_j(x,L)=\{\lambda_i^{(j)}\in[x,x+L)\}$. The number variance for an individual molecular geometry is then calculated as
\begin{align}
\Sigma_j^2(L)=
\left\langle
\left[\mathcal{N}_j(x,L)-L\right]^2
\right\rangle_x,
\end{align}
where the average is performed over different starting positions $x$ \footnote{To reduce finite-size effects associated with the spectral boundaries, the first and last five unfolded energy units were excluded when selecting the interval positions.}. For each window length $L$, 150 equal subdivisions were used to generate uniformly spaced starting positions within the allowed spectral region. The ensemble-averaged number variance was then obtained by averaging the individual number variances over all molecular geometries: $\Sigma^2(L) = \left\langle \Sigma_j^2(L) \right\rangle_j$. This procedure preserves the individual spectral boundaries and avoids combining energy levels belonging to different molecular geometries during the interval-counting process.

\begin{figure}[!htbp]
    \centering
    
    \begin{subfigure}[b]{0.4\textwidth}
        \centering
        \includegraphics[width=\textwidth]{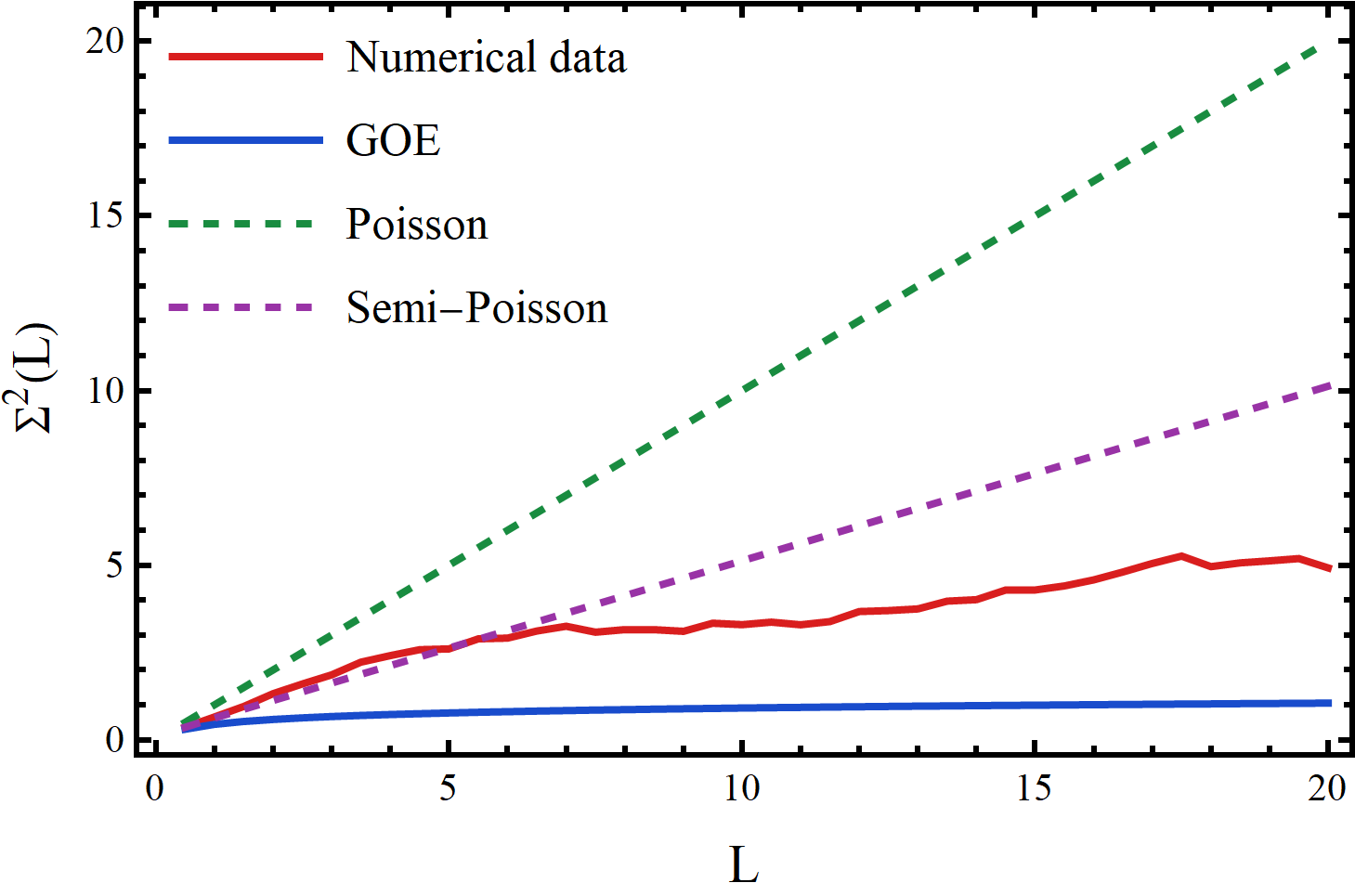}
        \subcaption*{(a) Number variance for AT.}
    \end{subfigure}
    
    \vspace{1em}
    
        \begin{subfigure}[b]{0.4\textwidth}
        \centering
        \includegraphics[width=\textwidth]{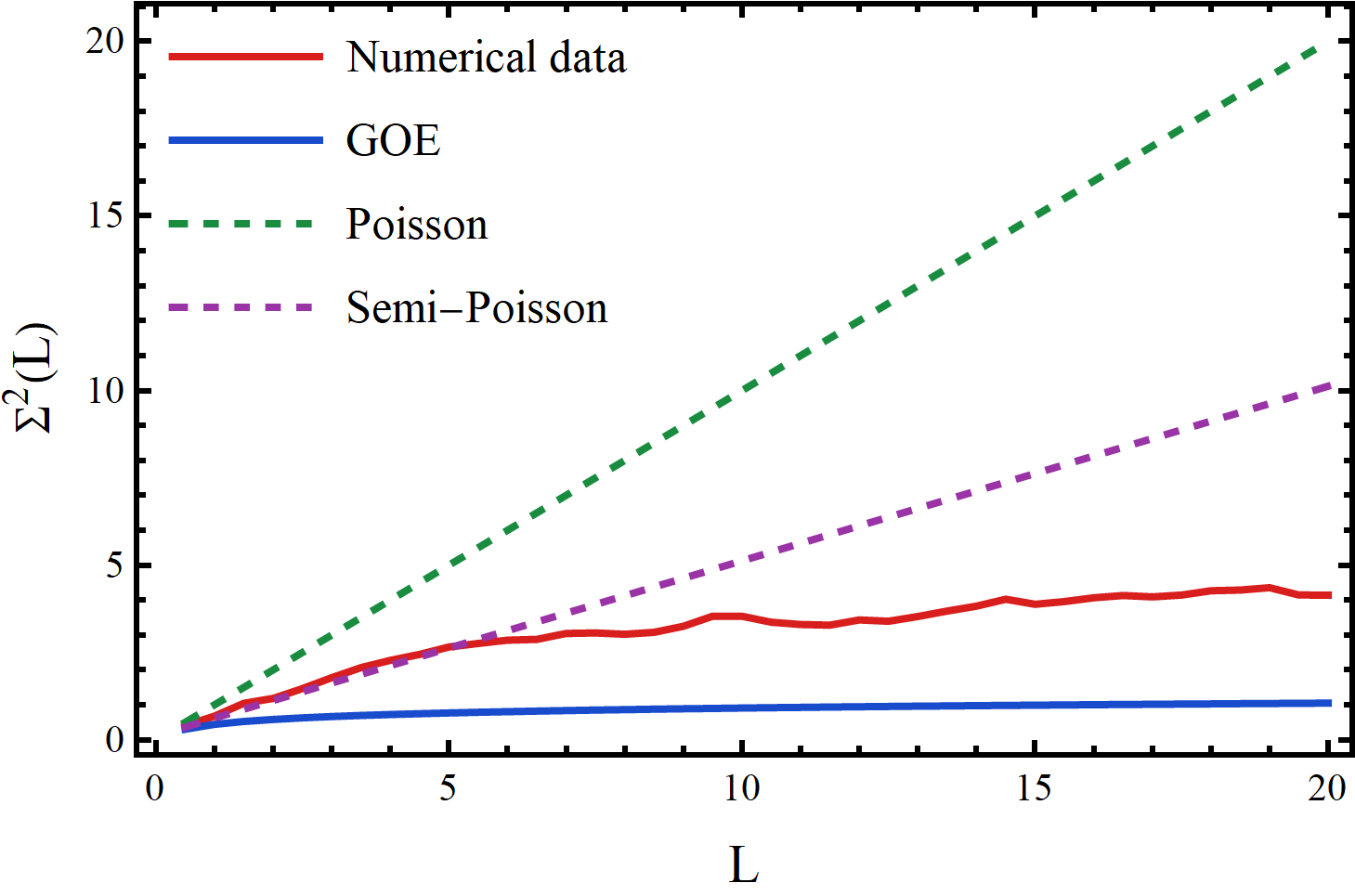}
        \subcaption*{(b) Number variance for GC.} 
    \end{subfigure}
    
    \vspace{1em}
    
    \begin{subfigure}[b]{0.4\textwidth}
        \centering
        \includegraphics[width=\textwidth]{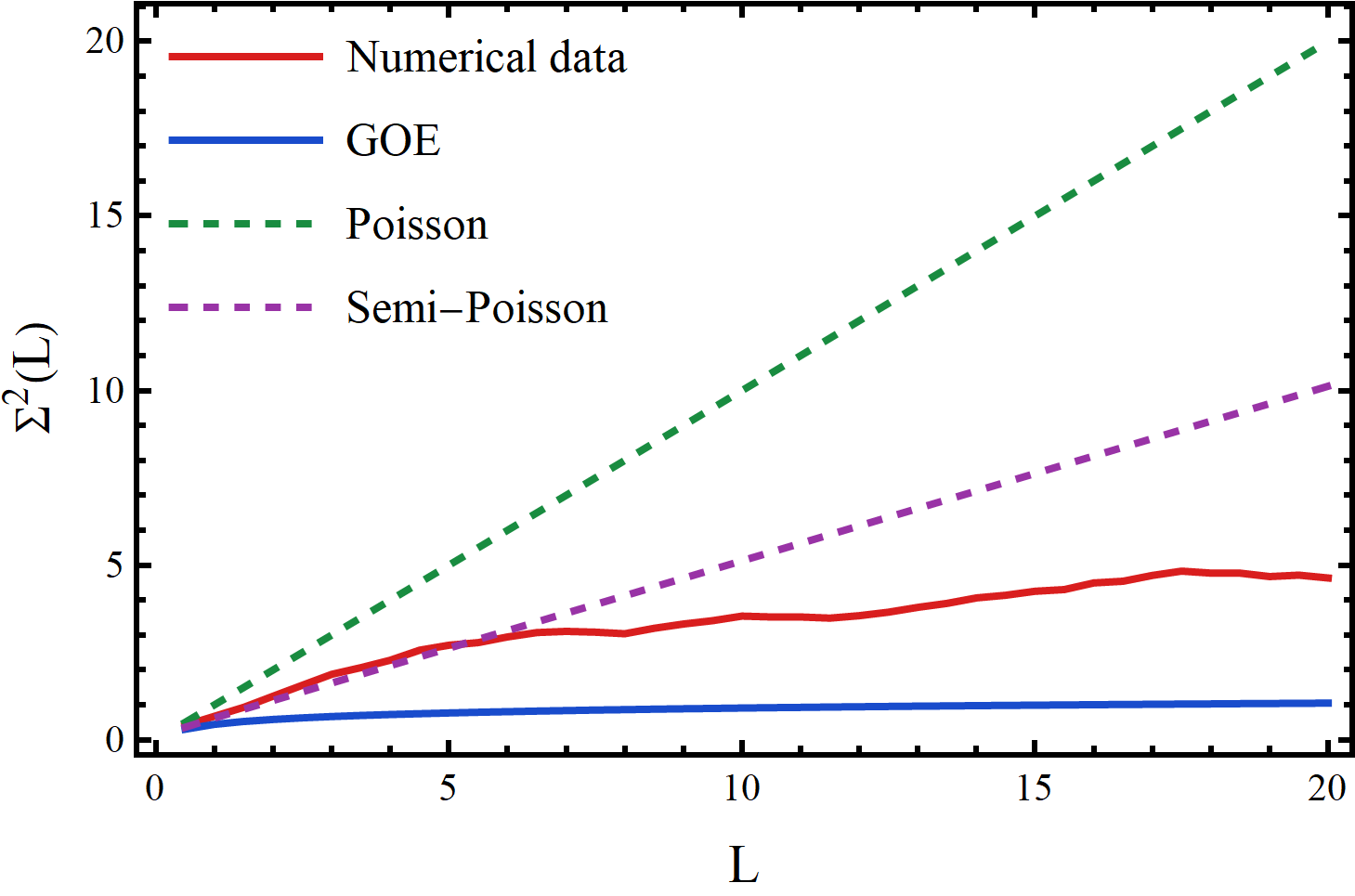}
        \subcaption*{(c) Number variance for for both bps.}
    \end{subfigure}
    \hspace{0.3cm} 

    \caption{The ensemble-averaged number variance for AT, GC, and combined base-pair datasets is plotted for the HB variation ensemble. The theoretical predictions for Poisson, SP, and GOE statistics are depicted and compared with the numerical results derived from the orbital energy spectra of the HB ensemble.}
    \label{fig:NV-HB}
\end{figure}

The number variance was evaluated for interval length range $L=\{0.5, 20\}$, in increments of $\Delta L=0.5$. Finally, to characterize the long-range spectral correlations, the calculated number variance was compared with the theoretical predictions of the Poisson, GOE, and SP statistics, as shown in fig.\eqref{fig:NV-HB}.

For all ensembles generated via hydrogen-bond length variations, the calculated number variance shows tight agreement with the semi-Poisson prediction for short-to-intermediate window lengths up to $L \approx 5$. For larger windows, the curve deviate downward from the SP baseline, settling into an intermediate regime bounded between the linear growth of SP and the logarithmic growth of the GOE statistics.

As reported in table.\eqref{tab:AGR-Chi}, the average compressibility for the three ensembles is $\langle\chi\rangle=0.1444$, that confirms the presence of robust long-range level correlations that enhance spectral rigidity beyond semi-Poisson expectations while remaining weaker than full GOE chaos.

\begin{table}[ht]
\centering
\caption{Mean AGR and compressibilities.}
\label{tab:AGR-Chi}

\small
\renewcommand{\arraystretch}{1.3}
\setlength{\tabcolsep}{10pt}

\resizebox{\columnwidth}{!}{%
\begin{tabular}{|c|c|c|c|c|}
\hline

\rowcolor{header}
\color{white}\textbf{ensemble} &
\color{white}\textbf{base pairs} &
\color{white}\textbf{mean AGR} &
\color{white}\textbf{Compressibility} \\
\hline

\multirow{3}{*}{HB}
& AT  & 0.46362 & 0.1764 \\
& GC  & 0.467297 & 0.1094 \\
& both bps  & 0.465444 & 0.1474 \\
\hline

\multirow{3}{*}{\cellcolor{lightrow}}
& \cellcolor{lightrow} AT
& \cellcolor{lightrow} 0.46553
& \cellcolor{lightrow} 0.1631 \\
\cline{2-5}

\cellcolor{lightrow}{MM}
& \cellcolor{lightrow} GC
& \cellcolor{lightrow} 0.477781
& \cellcolor{lightrow} 0.1099 \\
\cline{2-5}

\cellcolor{lightrow}
& \cellcolor{lightrow} both bps
& \cellcolor{lightrow} 0.471607
& \cellcolor{lightrow} 0.1478 \\
\hline

\cellcolor{LightOrange}{GOE}
& \cellcolor{LightOrange} -
& \cellcolor{LightOrange} 0.536
& \cellcolor{LightOrange} 0 \\
\hline

\cellcolor{LightOrange}{SP}
& \cellcolor{LightOrange} -
& \cellcolor{LightOrange} 0.500
& \cellcolor{LightOrange} 0.5 \\
\hline

\cellcolor{LightOrange}{Poisson}
& \cellcolor{LightOrange} -
& \cellcolor{LightOrange} 0.386
& \cellcolor{LightOrange} 1 \\
\hline

\end{tabular}%
}
\end{table}

In the next section, we move beyond single-variable hydrogen-bond shifts to consider the multi-mode variation protocol. Here, we analyze distorted base-pair geometries where the constituent atoms are no longer treated as rigid units. Instead, all atomic coordinates are permitted to fluctuate independently in all spatial directions around their equilibrium positions. This multi-mode structural ensemble allows us to evaluate how internal vibrational modes and intra-monomer geometric deformations influence orbital level repulsion and spectral rigidity compared to pure hydrogen-bond stretching.

\section{Multimode variation\label{MM}}

All constituent atoms of the base pairs are allowed to translate freely in three-dimensional space. This translations can originates from an external shock wave - a perturbative effect coming from thermal or non-thermal distortion.  This is considered as a refinement over the rigid HB scan, which permitted only relative shifts between rigid monomers. By allowing full three-dimensional atomic displacements, the multimode ensemble captures a more physically realistic representation of molecular dynamics and environmental perturbations.
\\
Rather than freezing intramolecular geometries, we apply random three-dimensional coordinate displacements $\{\Delta x, \Delta y, \Delta z\}$ to all atoms of the optimized Watson-Crick AT and GC base-pair structures, permitting both bond lengths and bond angles to vary across realizations. The sampling is constrained so that the native hydrogen-bonding is preserved - maintaining two hydrogen bonds in AT and three in GC - and we generate an ensemble of 1000 multimode geometries for each base pair. Each distorted configuration can thus be interpreted as an instantaneous snapshot along a perturbed dynamical trajectory.

Across all generated geometries, the Kohn--Sham orbital energies span from $-520.548\,\mathrm{eV}$ to $105.058\,\mathrm{eV}$ for AT, and from $-520.555\,\mathrm{eV}$ to $104.59\,\mathrm{eV}$ for GC. After unfolding and normalization, the corresponding minimum and maximum level spacings $\{\min(s), \max(s)\}$ are $\{1.01\times 10^{-3},\, 13.4328\}$ for AT and $\{1.01\times 10^{-3},\, 10.3224\}$ for GC.

\begin{figure}[!htbp]
    \centering
    
    \begin{subfigure}[b]{0.4\textwidth}
        \centering
        \includegraphics[width=\textwidth]{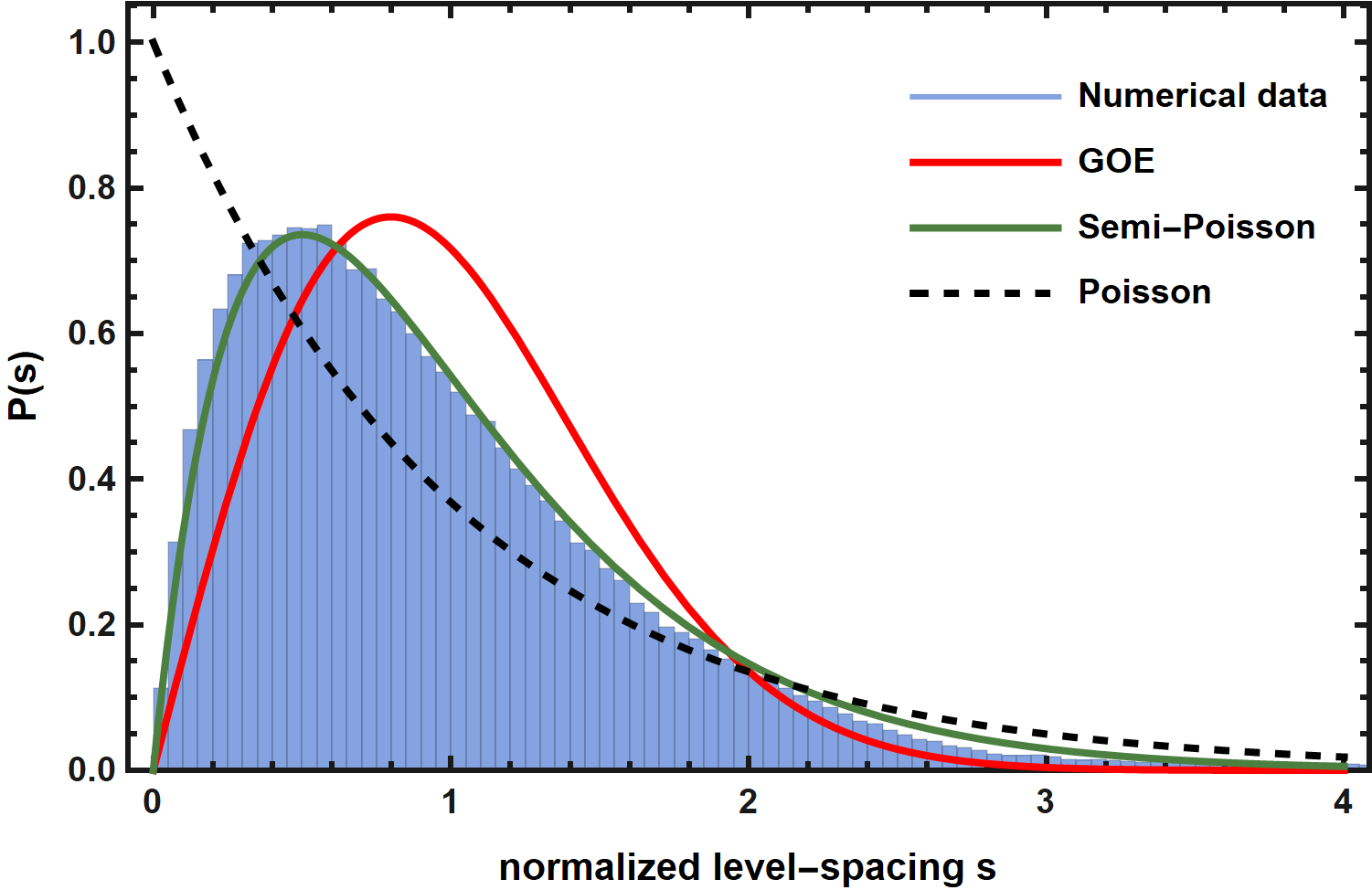}
        \subcaption*{(a) level-spacings of orbital energies for AT.}
    \end{subfigure}
    
    \vspace{1em}
    
        \begin{subfigure}[b]{0.4\textwidth}
        \centering
        \includegraphics[width=\textwidth]{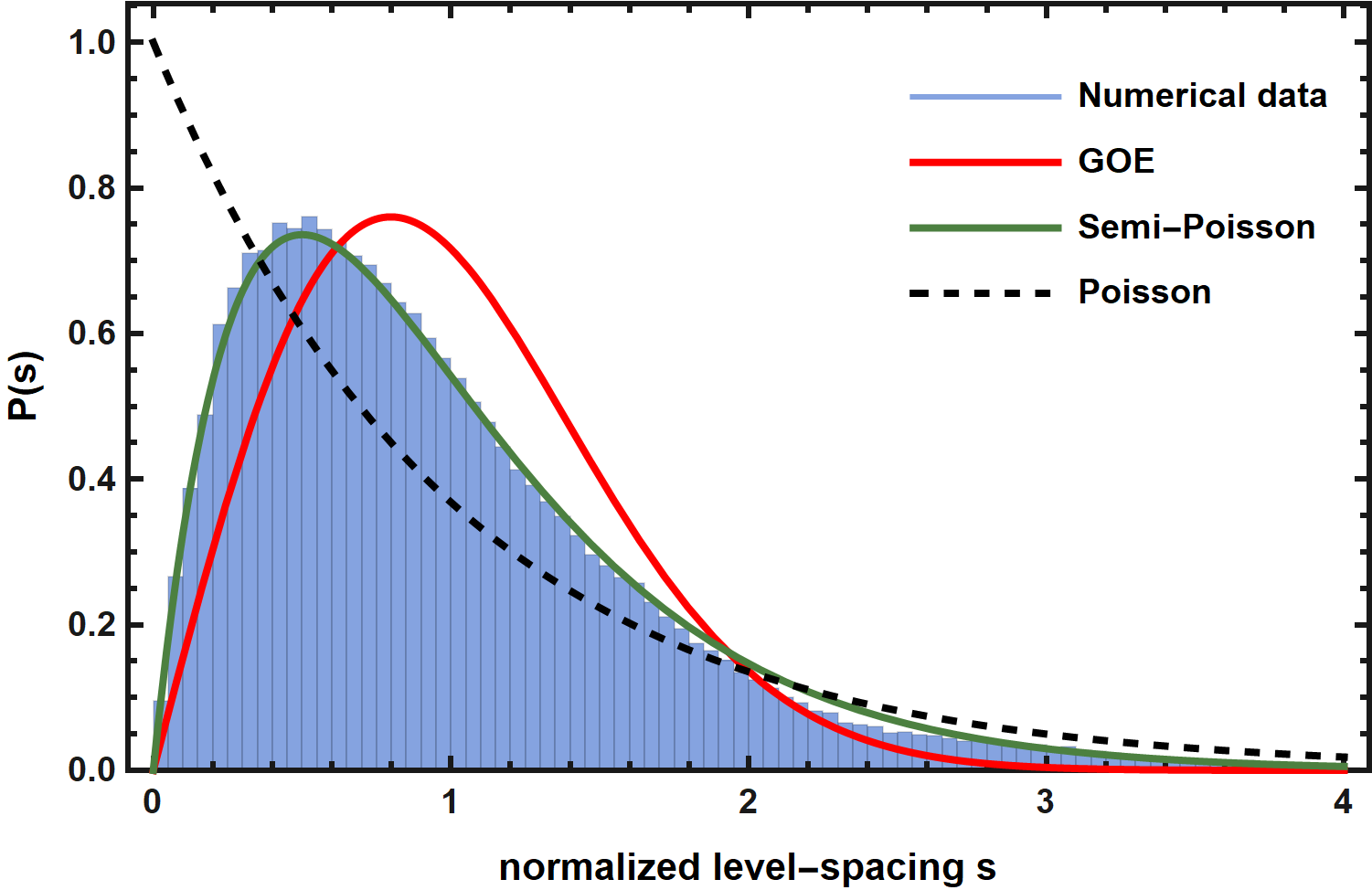}
        \subcaption*{(b) level-spacings of orbital energies for GC.}
    \end{subfigure}
    
    \vspace{1em}
    
    \begin{subfigure}[b]{0.4\textwidth}
        \centering
        \includegraphics[width=\textwidth]{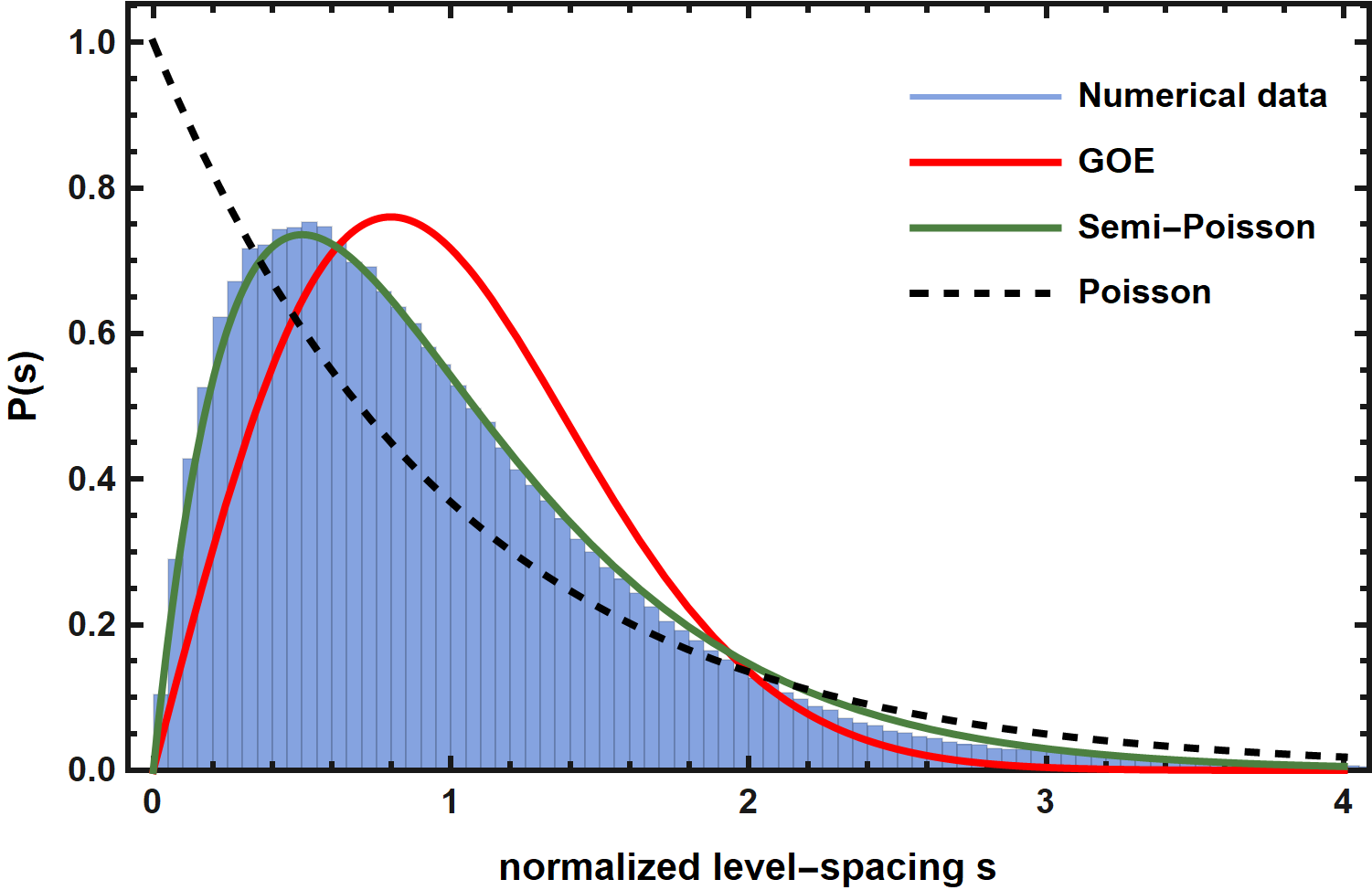}
        \subcaption*{(c) level-spacings of orbital energies for both bps.}
    \end{subfigure}
    \hspace{0.3cm} 

    \caption{The probability distribution of normalized level-spacing of orbital energies for MM variation with bin width of $\Delta s=0.05$ is plotted for AT, GC and combined base pairs. The comparison is made with theoretical curves, demonstrating exceptionally strong agreement with SP statistics.}
    \label{fig:OE-MM}
\end{figure}

Repeating the level-spacing analysis on the orbital energy levels of this new ensemble allows us to test whether full three-dimensional structural variations induce signatures of quantum chaos. Remarkably, the resulting distributions match the semi-Poisson distribution exceptionally well, as shown in fig.\eqref{fig:OE-MM}.

\begin{figure}[!htbp]
    \centering
    
    \begin{subfigure}[b]{0.4\textwidth}
        \centering
        \includegraphics[width=\textwidth]{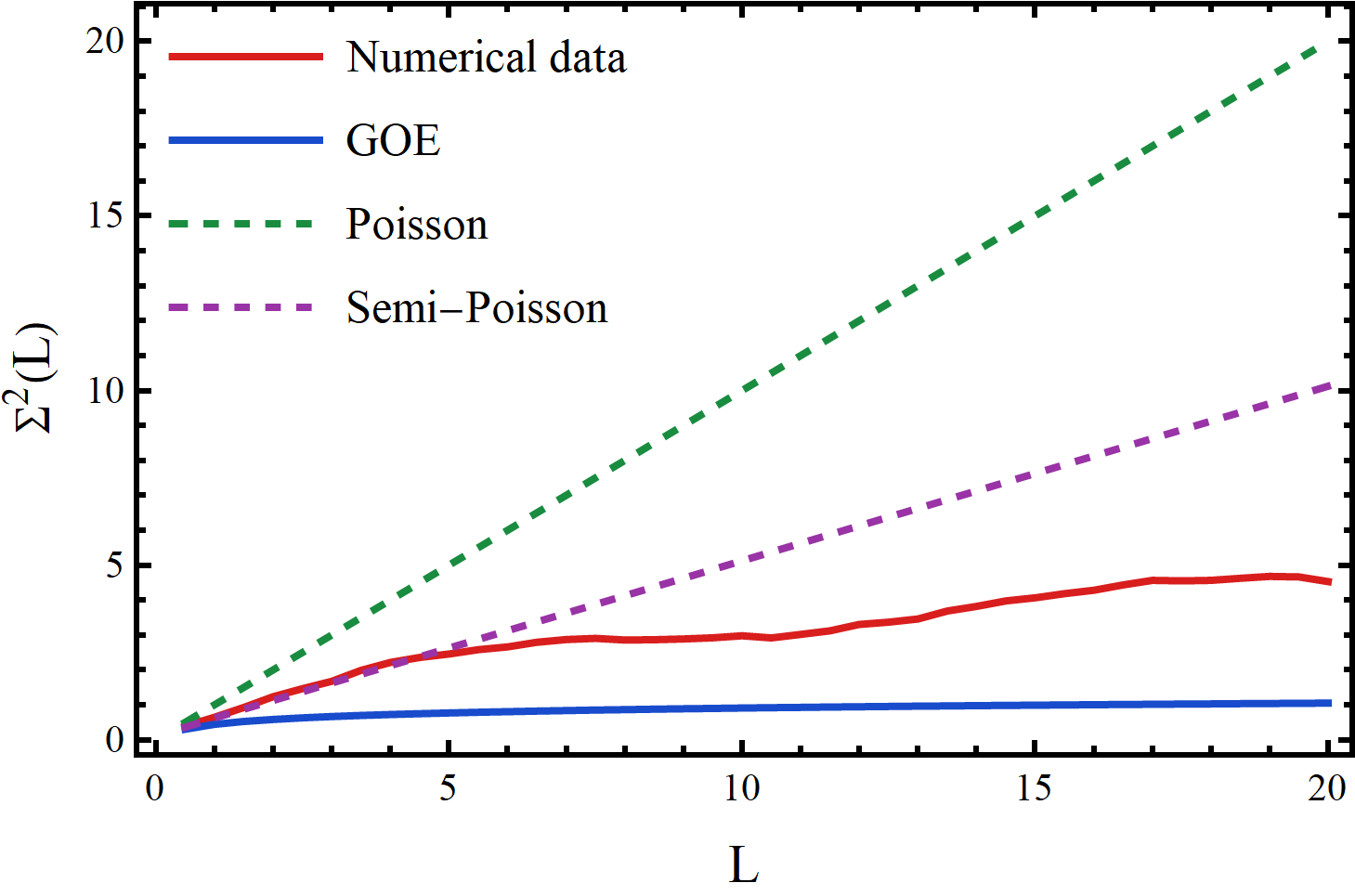}
        \subcaption*{(a) Number variance for AT.}
    \end{subfigure}
    
    \vspace{1em}
    
        \begin{subfigure}[b]{0.4\textwidth}
        \centering
        \includegraphics[width=\textwidth]{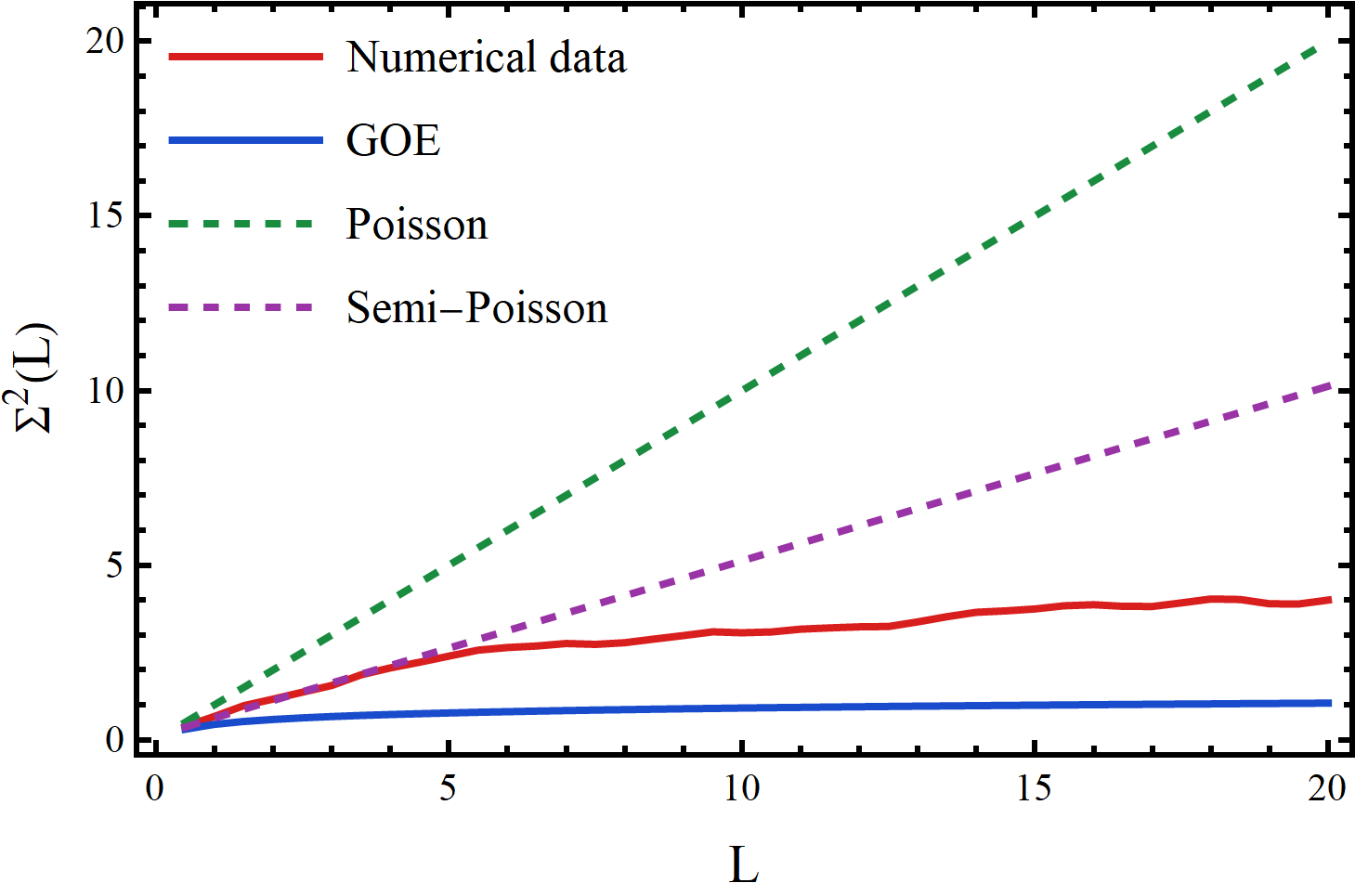}
        \subcaption*{(b) Number variance for GC.}
    \end{subfigure}
    
    \vspace{1em}
    
    \begin{subfigure}[b]{0.4\textwidth}
        \centering
        \includegraphics[width=\textwidth]{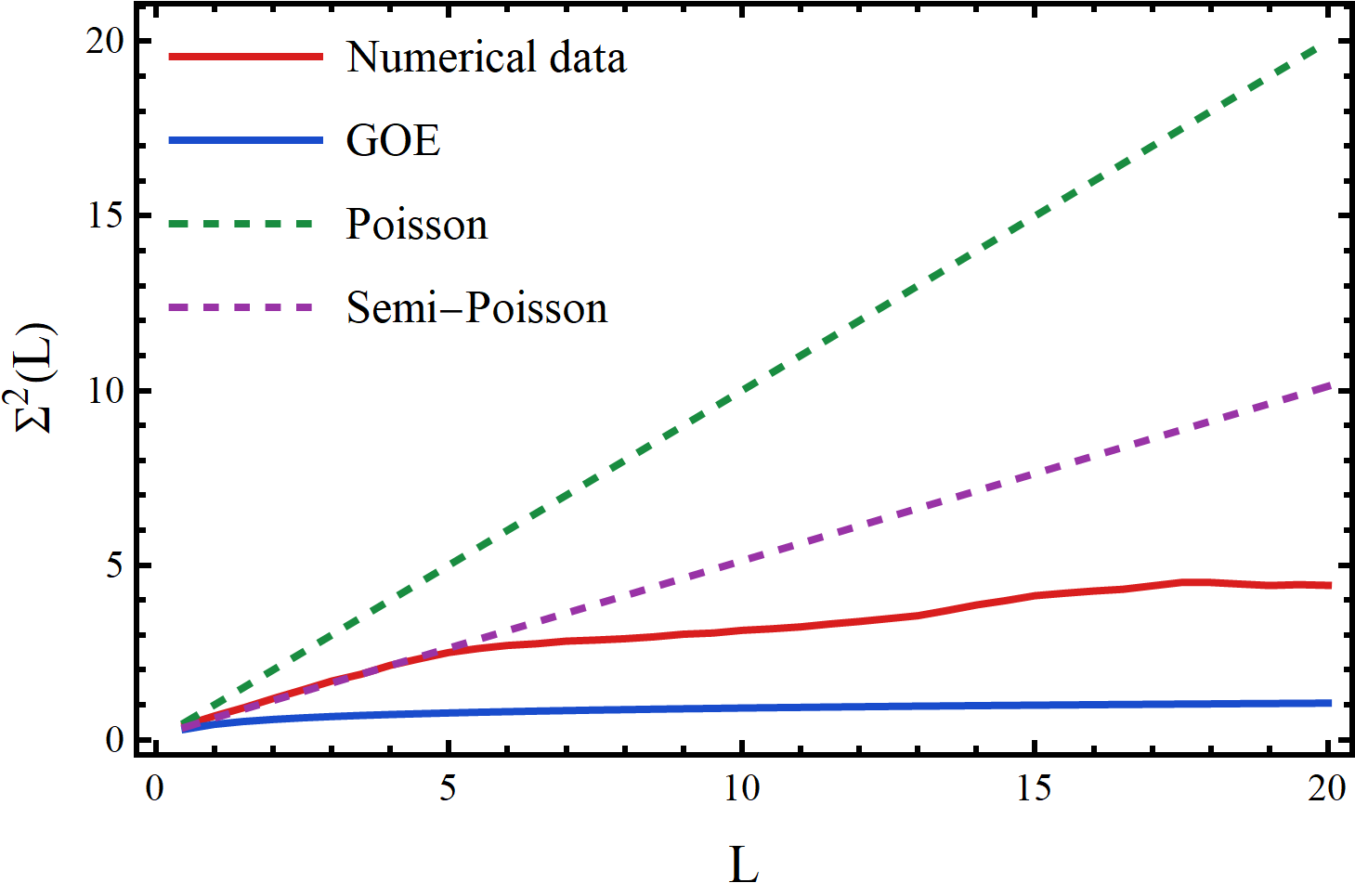}
        \subcaption*{(c) Number variance for both bps.}
    \end{subfigure}
    \hspace{0.3cm} 

    \caption{The number variance for three base-pairing cases is plotted for the MM variation ensemble.}
    \label{fig:NV-MM}
\end{figure}

For this case, we obtain average values of the adjacent-gap ratio and spectral compressibility across the three base-pairing ensembles of $\langle r \rangle = 0.471$ and $\langle \chi \rangle = 0.1402$, respectively. These values point to the same regime found in the previous section, i.e. short-range level repulsion accompanied by strong long-range spectral correlations. Nearly the same number-variance behavior also emerges for the multimode variation ensemble (see Fig.~\ref{fig:NV-MM}).

Taken together, the spectral statistics of both the hydrogen-bond and multimode ensembles demonstrate that while structural fluctuations perturb the system away from integrability, short-range repulsion remains robustly governed by semi-Poisson statistics with intermediate long-range correlation. Physical origins and broader implications of this finding are evaluated in the following section.

\section{Discussion\label{Dis}}

In this work, we investigated the statistical properties of electronic energy levels in ensembles of Watson--Crick AT and GC base pairs within the framework of random matrix theory. By analyzing the unfolded energy spectra, we probed both short-range and long-range spectral correlations using nearest-neighbor level-spacing distributions, adjacent-gap ratios, the number variance, and spectral compressibility. This analysis was carried out on structural ensembles generated by distorting the optimized geometries through hydrogen-bond and multimode variations.

\vspace{0.5cm}

A notable finding is that for the multimode-variation ensemble, the level-spacing distribution closely matches the semi-Poisson distribution. Furthermore, level repulsion and strong long-range correlations are observed across the orbital energy spectra of the base pairs.

\vspace{0.5cm}

The observed semi-Poisson statistics may suggest that the electronic structure of DNA exhibits features associated with {\it pseudo-symmetry} \cite{psudosym} rather than conventional symmetry structures, an interesting possibility that warrants further study.

\vspace{0.2cm}

Although the main analysis focuses on bulk level spacings, we also observe that near the extreme boundaries of the single-particle spectrum, the distribution of extreme eigenvalues transitions from bulk statistics to Tracy-Widom statistics~\cite{TW1,TW2,TW3}. This contrast between bulk and edge statistics motivates the following biophysical implications and open hypotheses.

\vspace{0.2cm}

These findings may have implications for understanding the electronic properties of DNA, particularly charge transport and charge-transfer processes \cite{ctransfer1,ctransfer2,ctransfer3,ctransfer4}. The intermediate spectral correlations may reflect the interplay between electronic level repulsion and structural disorder in DNA molecules. Furthermore, the Tracy-Widom behavior of the spectral edges provides a statistical description of fluctuations in the extreme electronic energy levels, potentially relevant to electron and hole transfer processes.

\vspace{0.2cm}

Since electronic charge transfer in DNA can contribute to oxidative damage \cite{damage} (e.g. for guanine) and radical formation, the observed spectral correlations may provide a statistical perspective on electronic processes potentially involved in mutation susceptibility. In particular, differences between AT and GC base pairs may influence their electronic properties and charge-transfer behavior. However, establishing a direct relationship between the observed spectral statistics and mutation rates requires further investigation.

\vspace{0.2cm}

Another interesting observation is that the semi-Poisson statistics predominantly emerge from the virtual orbital spectrum, suggesting that unoccupied electronic states exhibit distinctive spectral correlations that may be relevant to understanding their role in DNA electronic properties and charge-transfer.

\vspace{0.2cm}

Future research could investigate a possible coupling-driven localization-delocalization transition in DNA using a tight-binding Hamiltonian with variable inter-base coupling \cite{tight-binding}. By examining the evolution of spectral statistics and eigenstate localization, one could identify a possible {\it critical coupling} and explore its implications for electronic transport in DNA.

\vspace{0.2cm}

Several other directions can be pursued in future work. First, a systematic comparison of AT and GC systems, as well as larger DNA sequences, could help establish whether spectral statistics depend on nucleotide composition and sequence organization. Second, investigating the effects of molecular geometry, temperature, solvent environment, and structural fluctuations could clarify the robustness of the observed statistical behavior. Furthermore, extending the analysis to electronic coupling, charge-transfer states, and excited-state properties may reveal possible connections between spectral correlations and charge transport in DNA.

\vspace{0.2cm}

One could also apply the same analysis to alternative base-pair geometries, such as reversed Watson-Crick, Hoogsteen, and reversed Hoogsteen pairings, to see how level-spacing distributions and other statistical metrics respond.

\vspace{0.2cm}

More broadly, this framework points to an intriguing avenue for applying random matrix theory to questions in molecular biology and genetics, particularly in testing whether distinct sequences, mutations, or structural modifications leave distinguishable signatures in electronic spectra.


\section*{Acknowledgement}
The author would like to express special gratitude to Hossein Babaei Aghbolagh for granting access to a high-performance computing system, without which these calculations could not have been completed. Special thanks are also extended to Mehdi Mosaiean for assistance with technical Python coding, and to Pavel Hobza, Ahmad Asoodeh, Saba Torkmann, and Hadis Nabavi for valuable discussions.




\end{document}